\documentclass[11pt,a4paper]{article}
\pdfoutput=1
\usepackage{jheppub}

\usepackage{ifpdf}

\usepackage{afterpage}

\usepackage{amssymb}
\usepackage{amsmath}
\usepackage{graphicx}
\usepackage{longtable}
\usepackage{verbatim}
\usepackage{amsfonts}
\usepackage{bbold}
\usepackage[table,usenames,dvipsnames]{xcolor}
\usepackage{titlesec}

\usepackage{enumerate}

\newcommand{\bear}{\begin{array}}
\newcommand{\ear}{\end{array}}

\newcommand{\beq}{\begin{equation}}
\newcommand{\eeq}{\end{equation}}
\newcommand{\beqa}{\begin{eqnarray}}
\newcommand{\eeqa}{\end{eqnarray}}

\newcommand{\nn}{\nonumber}
\def\OMIT#1{{}}
\newcommand{\lsim}{\mathrel{\rlap{\lower4pt\hbox{\hskip1pt$\sim$}}
    \raise1pt\hbox{$<$}}}         
\newcommand{\gsim}{\mathrel{\rlap{\lower4pt\hbox{\hskip1pt$\sim$}}
    \raise1pt\hbox{$>$}}}         

\usepackage{tikz}
\usetikzlibrary{trees}
\usetikzlibrary{decorations.pathmorphing}
\usetikzlibrary{decorations.markings}
\tikzset{
    photon/.style={decorate, decoration={snake}, draw=black},
    wino/.style={draw=redwine},
    electron/.style={draw=black, postaction={decorate},
        decoration={markings,mark=at position .55 with {\arrow[draw=black]{>}}}},
    scalar/.style={draw=black, dashed,postaction={decorate},
        decoration={markings,mark=at position .55 with {\arrow[draw=black]{>}}}},
    gluon/.style={decorate, draw=black,
        decoration={coil,amplitude=4pt, segment length=5pt}}
}

\newcommand{\ev}{\mathrm{eV}}

\newcommand{\Mev}{\mathrm{MeV}}
\newcommand{\Gev}{\mathrm{GeV}}
\newcommand{\Tev}{\mathrm{TeV}}

\newcommand{\Pl}{\mathrm{Pl}}
\newcommand{\SM}{\mathrm{SM}}
\newcommand{\QCD}{\mathrm{QCD}}

\newcommand{\uv}{\mathrm{UV}}
\newcommand{\ir}{\mathrm{IR}}

\newcommand{\ini}{{\mathrm{ini}}}
\newcommand{\aft}{{\mathrm{after}}}
\newcommand{\dil}{{\mathrm{dil}}}
\newcommand{\mis}{{\mathrm{mis}}}

\newcommand{\mO}{\ensuremath{\mathcal{O}}}
\newcommand{\mL}{\ensuremath{\mathcal{L}}}
\newcommand{\mV}{\ensuremath{\mathcal{V}}}

\newcommand{\Sec}[1]{Sec.~\ref{#1}}

\newcommand{\Fig}[1]{Fig.~\ref{#1}}

\newcommand{\Eq}[1]{Eq.~(\ref{#1})}
\newcommand{\Eqs}[2]{Eqs.~(\ref{#1}) and (\ref{#2})}
\newcommand{\Eqst}[2]{Eqs.~(\ref{#1})-(\ref{#2})}
\newcommand{\eg}{\textit{e.g.}\ }
\newcommand{\ie}{\textit{i.e.}\ }

\newcommand{\bl}{\left}
\newcommand{\br}{\right}

\newcommand{\ignore}[1]{}

\vspace*{-30mm}

\title{\boldmath Dynamical axion misalignment with small instantons}
\author[a]{Manuel A. Buen-Abad,}
\author[b]{JiJi Fan}
\affiliation[a]{Department of Physics, Brown University, Providence, RI, 02912, USA}
\affiliation[b]{Department of Physics and Brown Theoretical Physics Center, Brown University, Providence, RI, 02912, U.S.A.}
\emailAdd{manuel\_buen-abad@brown.edu}
\emailAdd{jiji\_fan@brown.edu}

\vspace*{1cm}

\abstract{
We present a new mechanism to relax the initial misalignment angle of the QCD axion and raise the cosmological bound on the axion decay constant. The QCD axion receives a contribution from small UV instantons during inflation, which raises its mass to the inflationary Hubble scale. This makes the axion start rolling down its potential early on. In the scenario, the standard model Yukawa couplings of quarks are dynamical, being of order one during the inflationary era and reducing to their standard model values once it ends. This means that after inflation the contribution of the small instantons is suppressed, and the axion potential reduces to the standard one from the usual IR instantons. As a result, when the axion starts to oscillate again after inflation, the initial misalignment angle is suppressed due to the dynamics during inflation. While the general idea of dynamical axion misalignment has been discussed in the literature before, we present in detail the major bottleneck on the mismatching between the minima of the axion potentials during and after inflation, and how it is circumvented in our scenario via the Froggatt-Nielsen mechanism. Taking into account of all the constraints, we find that the axion decay constant could be raised to the GUT scale, $10^{15}~\Gev$, in our scenario.
}

\begin{document}

\maketitle



\section{Introduction}
\label{sec:intro}

The QCD axion is a leading solution to the strong CP problem in the standard model~\cite{Peccei:1977hh, Peccei:1977ur, Weinberg:1977ma, Wilczek:1977pj, Kim:1979if, Shifman:1979if, Zhitnitsky:1980tq, Dine:1981rt}. In the solution, there is a global $U(1)$ Peccei-Quinn (PQ) symmetry, which has a mixed anomaly with the QCD gauge group $SU(3)_c$. $U(1)_{PQ}$ is broken spontaneously in the UV at a high energy scale $f_a$, resulting in a pseudo-Nambu-Goldstone boson, the axion. The axion makes the $\theta$ angle of QCD dynamical. Non-perturbative QCD effects generate a periodic potential for the axion, at the minimum of which the strong CP phase is relaxed to be zero. Phenomenologically, the zero-temperature axion potential could be written as\footnote{Note that the full potential is more complicated involving mixings between axion and pions~\cite{DiVecchia:1980yfw}. Yet for our discussion, the phenomenological potential is sufficient.}
\begin{equation}
  \Lambda_0^4 \left(1- \cos\left(\frac{a}{f_a}\right)\right),
\end{equation}
 where $\Lambda_0$ is given by \cite{Weinberg:1977ma}
\begin{equation}
  \Lambda_0^2 = \sqrt{\frac{m_um_d}{(m_u+m_d)^2}} f_\pi m_\pi \approx (75.5 \, {\rm MeV})^2,
\label{eq:IR}
\end{equation}
where $m_u, m_d$ are up and down quark masses and $m_\pi, f_\pi$ are the mass and decay constant of the pion, respectively.
The axion then obtains a mass (at zero temperature):
\begin{equation}
  m^\ir = \frac{\Lambda_0^2}{f_a}.
\label{eq:todaymass}
\end{equation}

Another attractive point of the QCD axion is that it could be a viable cold dark matter candidate. Its relic abundance could be generated through the misalignment mechanism~\cite{Preskill:1982cy, Dine:1982ah, Abbott:1982af}. In the simplest scenario, PQ symmetry breaking happens during inflation. Initially Hubble friction holds the axion at a random place in its field space with an initial misalignment angle $\theta_0$. Without fine-tuning, $\theta_0$ is expected to be $\sim \mO(1)$. When the Hubble scale drops around the axion mass, the axion starts to oscillate coherently around the minimum of its potential. The coherent oscillation redshifts as non-relativistic matter. It is found that when $f_a \sim 10^{12}~\Gev$, and assuming $\theta_0 \sim \mO(1)$, the relic abundance of the QCD axion matches the observed value of dark matter abundance. For a higher PQ breaking scale, axions would overclose the Universe. Some recent reviews on axion cosmology can be found in Refs.~\cite{Sikivie:2006ni, Marsh:2015xka, Hook:2018dlk}.

From the top-down point of view, axions from many scenarios of string theory could have a high PQ breaking scale, above the cosmological upper bound. For instance, in models with GUT-like phenomenologies, $f_a$ could be around the GUT scale~\cite{Svrcek:2006yi}. While it is certainly possible to construct string-based models with $f_a$ much below the GUT scale, it is interesting to explore whether there is any cosmological scenario beyond the minimal misalignment model that could raise the upper bound on $f_a$ to make the GUT scale axions viable. There have already been quite a few different types of proposals in the literature. Examples include a tuned tiny initial misalignment angle with possibly an anthropic reason~\cite{Tegmark:2005dy}, late time entropy production before BBN to dilute the axion relic abundance~\cite{Kawasaki:1995vt, Banks:1996ea}, and transferring axion energy density to other species such as dark photons through particle production~\cite{Agrawal:2017eqm, Kitajima:2017peg}.

In this article, we focus on the possibility that the initial misalignment angle is dynamically relaxed to a small value when the QCD axion starts to oscillate. One way to achieve this is to make the axion much more massive during inflation, \eg from a much stronger QCD in the early Universe~\cite{Dvali:1995ce}. It has been argued in Ref.~\cite{Choi:1996fs} that it is challenging to realize this possibility in supersymmetric scenarios with several assumptions about relevant parameters. Yet recently Ref.~\cite{Co:2018phi} revisited this possibility and showed that if the Hubble induced Higgs mass squared parameter is negative and the Higgs field takes a large value along the flat direction in the supersymmetric scenario, one could make QCD strong enough to raise the axion mass to about or above the inflationary Hubble scale. One major challenge in this scenario is whether the minima of the potentials during and after inflation are sufficiently close to each other. This usually requires the assumption of an approximate CP symmetry with a tiny breaking~\cite{Banks:1996ea}. Another possible way to relax the initial misalignment angle is to have an exponentially long inflationary period as suggested in~\cite{Graham:2018jyp, Guth:2018hsa}.

We will explore a different mechanism, using small UV instantons~\cite{Agrawal:2017ksf}, to make the QCD axion heavy during inflation in order to relax the misalignment angle. Our main motivation is that the UV instanton model proposed in Ref.~\cite{Agrawal:2017ksf} does not introduce new CP phases beyond the standard model by construction. This could potentially address the main bottleneck of the dynamical misalignment angle scenario. Yet as every experienced model builder knows, life is never that easy. In order to apply this mechanism and relax the upper bound on $f_a$, we need to introduce additional modules to dynamically vary the axion mass during and after inflation. We employ dynamical Yukawa couplings in the Froggatt-Nielsen scenario to achieve this, and argue that the introduced CP phase is small enough not to spoil the mechanism. Combining the small instantons and dynamical Yukawa couplings, we construct a model in which the upper bound on $f_a$ could be raised up to the the GUT scale, $10^{15}~\Gev$.

The paper is organized as follows. In Sec.~\ref{sec:mechanism}, we outline the basic idea of the dynamical misalignment mechanism and the minimum requirements to make it work. In Sec.~\ref{sec:model}, we present the two main ingredients of our model: the mechanism based on small instantons to raise the axion mass during inflation, and the dynamical Yukawa coupling mechanism to relax the axion mass after inflation. We identify the allowed parameter space satisfying all the requirements, and compute the relic abundance of the QCD axion in this scenario. In Sec.~\ref{sec:other}, we discuss another possible variant of the model and comment on coincident energy scales in our model. We conclude in Sec.~\ref{sec:conclusion}.

\section{The basic mechanism and requirements}
\label{sec:mechanism}

We start with explaining our notations. In our model some quantities are dynamical, \ie they have different values during and after inflation. In order to avoid a cumbersome clustering of indices, we use a tilde ($\sim$) above to denote the values of these quantities {\it during} inflation; and no tildes for their values {\it after} it. For example, the mass of the QCD axion is $\tilde{m}$ during the inflationary era, and simply $m$ after it ends.

\begin{figure}
\begin{center}
\includegraphics[width=0.5\textwidth]{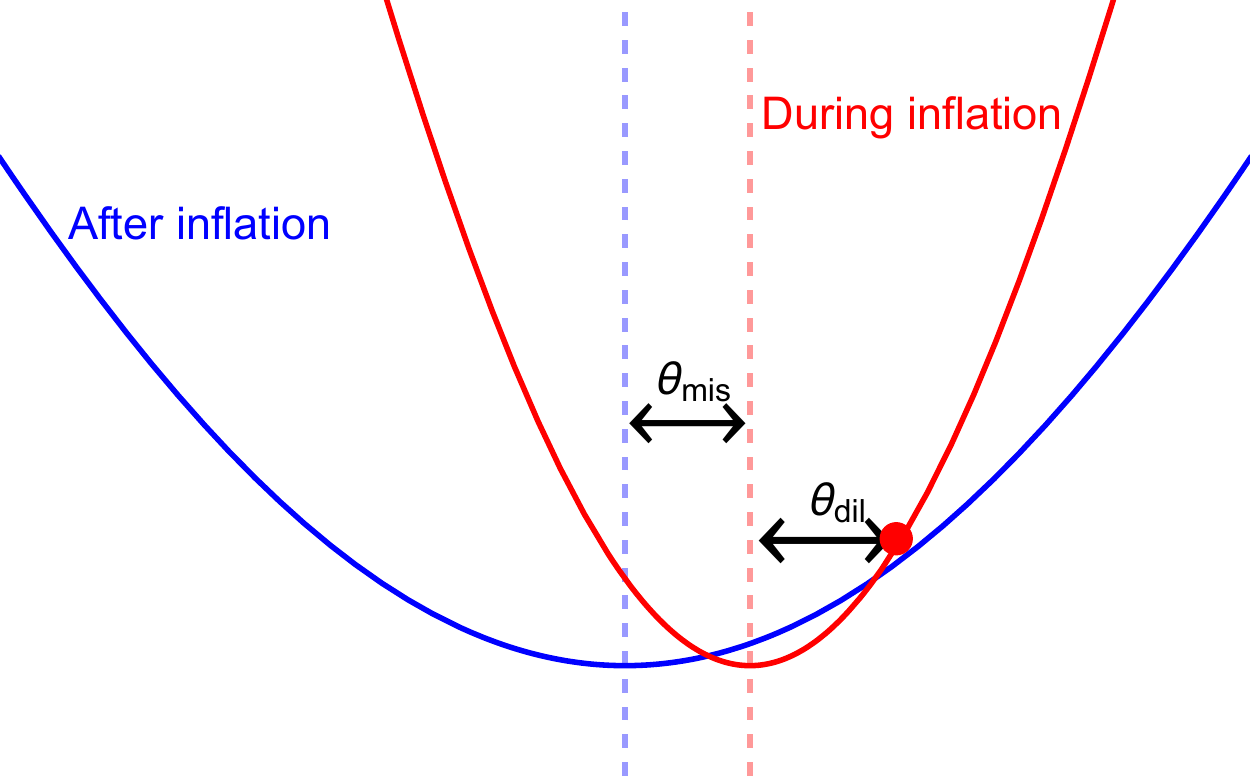}
\caption{A schematic picture of the scenario. Red: axion potential during inflation; blue: axion potential after inflation. The red dot represents the axion at the end of the inflation. Dashed vertical lines indicate the minima of the potentials.}
\label{fig:schematic}
\end{center}
\end{figure}

In this section we will now describe our basic scenario to relax the initial misalignment angle of the QCD axion, and discuss the main requirements to realize this scenario. During inflation, the mass of the QCD axion receives its dominant contribution from the UV instantons, which we denote by $\tilde{m}^\uv$, and is raised about the Hubble scale. Consequently, it starts to roll down its potential. The axion field value redshifts away exponentially and the misalignment angle, measured from the minimum of the axion potential {\it during inflation}, is suppressed at the end of the inflation. We will denote this diluted angle as $\theta_\dil$, which is the axion field value at the end of the inflation divided by its decay constant. After inflation, a mechanism (in our case, dynamical Yukawa couplings) reduces the UV instanton contribution, now $m^\uv$, to be negligible compared to the contribution from the usual IR instantons, $m^\ir$. Depending on the model, the minima of the axion potential during inflation and after inflation may not perfectly overlap with each other. In other words, there could be a mismatch angle, $\theta_\mis$, which is defined as the displacement between the two minima in the field space divided by the axion decay constant. This could be due to new CP phases present during inflation. The evolution of the axion potential after inflation becomes the same as the standard one in the canonical misalignment mechanism. When the Hubble scale drops to be around the axion mass after inflation, the axion starts to oscillate with an initial misalignment angle $\theta_\ini \sim {\rm max}[\theta_\dil, \theta_\mis]$. Since the relic abundance of the axion is proportional to $\theta_\ini^2$, it is suppressed at a given $f_a$ when ${\rm max}[\theta_\dil, \theta_\mis] \ll 1$. Equivalently, the cosmological upper bound on $f_a$ could be raised when $\theta_\ini$ is small. The schematic picture of the scenario is shown in Fig.~\ref{fig:schematic}.

It is clear that in order for the scenario to be feasible, one needs to satisfy at least the following four requirements:
\begin{itemize}
\item A large enough UV instanton's contribution to raise the axion mass around or above the Hubble scale during inflation: $\tilde{m}^\uv \gtrsim H_I \gg \tilde{m}^\ir$.
\item Enough dilution of the initial misalignment angle due to inflation: $\theta_{\rm dil} \ll 1$.
\item Approximate alignment of the minima of the QCD axion potential in the early Universe and today: $\theta_{\rm mis} \ll 1$. This requires any new CP phases that might have been introduced in the mechanism to be tiny.
\item Suppression of the UV instanton's contribution to the QCD axion after inflation, so that its mass reduces to the usual value determined by the IR instantons: $m^\uv \ll m^\ir$.
\end{itemize}
While we focus on using the small instantons to raise the axion mass during inflation, the general requirements above apply to any model that tries to have a dynamical axion mass to relax the initial misalignment angle and lift the upper bound on the axion decay constant. While it is possible to have a mechanism to satisfy one of the four conditions, \eg to make the axion heavy during inflation, it is generally quite challenging to have a coherent story to meet all of them at the same time.

\section{The model}
\label{sec:model}
In this section, we will present a plausible model and its main ingredients to satisfy all the requirements outlined in the previous section. We use an enhanced short-distance instanton effects to raise the axion mass during inflation and a dynamical Yukawa coupling mechanism to suppress the UV instantons after inflation. We demonstrate that there is viable parameter space in which the initial misalignment angle when QCD axion starts to oscillate is greatly suppressed.

\subsection{UV instanton to raise the axion mass}
\label{sec:uv}

It has been proposed quite a while back that if QCD becomes strong at high energies, the correction from small color instantons to the axion potential could be significant~\cite{Holdom:1982ex, Holdom:1985vx, Dine:1986bg, Flynn:1987rs, Choi:1988sy, Choi:1998ep}.
In this type of scenario such as the model in Ref.~\cite{Flynn:1987rs}, one needs to change the signs of QCD beta function coefficients twice: once to make the QCD gauge coupling run stronger towards the UV and once to make it asymptotically free at an even higher energy. To achieve the first flip of the sign, one usually needs to introduce either a large number of fermions in small representations of $SU(3)_c$ such as fundamental representations or a vector-like pair of fermions in large representations. The potential issue with the first approach is that one could introduce new CP phases due to mixing of the fermions, which shifts the minimum of the QCD axion potential. For the second approach which only requires a couple of additional matter fields with no new CP phase, it is difficult to embed them in a grand unified theory to make QCD asymptotically free eventually. While there is clearly no no-go theorem for this scenario, we will not pursue this direction in this article.

Recently a new way is proposed to increase the gauge coupling at high energy and thus the contribution from small instanton through higgsing a product gauge group~\cite{Agrawal:2017ksf}.\footnote{Similar models in color unified theory have been developed in Ref.~\cite{Gaillard:2018xgk}, while others have been studied in the context of lepton flavor universality violation in $B$-meson decays \cite{Fuentes-Martin:2019bue}.} The main advantage of this approach is that no new CP phase is introduced. Since it is the main mechanism we rely on to increase the axion mass during inflation, we briefly review the key idea below.
Consider that at an energy scale $M \gg~\Tev$, the QCD gauge group emerges from higgsing a product gauge group:
\begin{equation}
  SU(3)_1 \times SU(3)_2 \to SU(3)_c.
\end{equation}
The higgsing could be achieved by having a complex scalar, $\Sigma$, transforming as a bi-fundamental, $(3, \bar{3})$, under the product gauge group. This complex scalar has a gauge invariant potential~\cite{Bai:2017zhj}
\begin{equation}
  \mV(\Sigma)= -m_\Sigma^2 {\rm Tr} \left(\Sigma \Sigma^\dagger\right) + \frac{\lambda}{2} \left( {\rm Tr}\left(\Sigma \Sigma^\dagger\right)\right)^2 + \frac{\kappa}{2} {\rm Tr} \left((\Sigma \Sigma^\dagger)^2\right),
\label{eq:linkpotential}
\end{equation}
where $\lambda, \kappa$ are order one real numbers barring fine-tunings and are positive so that the potential is bounded from below.\footnote{For the potential in Eq.~\eqref{eq:linkpotential}, we gauge an extra $U(1)$ factor which forbids trilinear terms such as Det$\Sigma$ and charge the link field under the $U(1)$.} When $m_\Sigma^2 >0$, $\Sigma$ obtains a vacuum expectation value (VEV), $\langle \Sigma \rangle = \frac{f_\Sigma}{2}  \mathbb{1}_3$ with
\begin{equation}
  f_\Sigma = \frac{m_\Sigma}{\sqrt{\kappa + 3 \lambda}},
\end{equation}
and breaks the product of $SU(3)$'s to the diagonal gauge group, which is identified as $SU(3)_c$. The symmetry breaking scale $M$ is defined as
\begin{equation}
  M^2 = (g_1^2+g_2^2) f_\Sigma^2,
\end{equation}
where $g_1$ and $g_2$ are the gauge couplings of $SU(3)_1$ and $SU(3)_2$ respectively.
The gauge couplings, $g_1$ and $g_2$ are related to the standard model strong coupling $g_s$ as
\begin{equation}
  \frac{1}{\alpha_s(M)} =\frac{1}{\alpha_1(M)} + \frac{1}{\alpha_2(M)},
\label{eq:matching}
\end{equation}
where $\alpha=g^2/(4\pi)$. One could see that to satisfy this relation, $\alpha_1~(\alpha_2)$ has to be larger than $\alpha_s$, which is exactly what we want to increase the contribution of the small UV instantons.

Consider two spontaneously breaking PQ symmetries $U(1)_{\rm PQ}$'s with breaking scales $f_1$ and $f_2$. For simplicity, we assume that $f_1 = f_2 \equiv f_a \gg M$, the scale at which the product gauge group is broken down to the diagonal. We also assume that only the standard model quarks are charged under the gauge group and there are no other new fermions charged under either $SU(3)$. Thus by construction, this model doesn't introduce new CP phases. Suppose that $U(1)_1$ has an anomaly with $(SU(3)_1)^2$ and $U(1)_2$ has an anomaly with $(SU(3)_2)^2$. Then we have two axions coupled to $SU(3)_1$ and $SU(3)_2$ respectively
\begin{equation}
  \frac{\alpha_1}{8 \pi} \left(\frac{a_1}{f_a} - \bar{\theta}_1\right) G_1 \tilde{G}_1+ \frac{\alpha_2}{8 \pi} \left(\frac{a_2}{f_a} - \bar{\theta}_2\right) G_2 \tilde{G}_2,
\end{equation}
where $G_1$ ($G_2$) is the field strength of $SU(3)_1$ ($SU(3)_2$) and $\bar{\theta}_1 (\bar{\theta}_2)$ is the effective theta angle of $SU(3)_1$ ($SU(3)_2$). At $M$, the non-perturbative UV instantons generate potentials for both $a_1$ and $a_2$, which can be computed using the dilute gas approximation~\cite{tHooft:1976snw} when $\alpha_1, \alpha_2$ remain perturbative and below one. In addition, after the symmetry breaking, both $a_1$ and $a_2$ are coupled to standard model QCD. In the effective field theory (EFT) below $M$, we have
\begin{equation}
  \Lambda_1^4 \cos\left(\frac{a_1}{f_a} - \bar{\theta}_1 \right) +\Lambda_2^4 \cos\left(\frac{a_2}{f_a} - \bar{\theta}_2 \right) + \frac{\alpha_s}{8\pi} \left(\frac{a_1}{f_a} - \bar{\theta}_1+\frac{a_2}{f_a} - \bar{\theta}_2\right) G \tilde{G},
\end{equation}
where $G$ is the field strength of $SU(3)_c$. The first two terms from the small UV instantons guarantee that the low energy effective theta angle $\bar{\theta}_{\rm eff} = \frac{a_1 + a_2}{f_a} - (\bar{\theta}_1 + \bar{\theta}_2)$ is relaxed to zero.

The short-distance instanton contribution to the axion potential is estimated to be, via dilute instanton gas approximation,
\begin{eqnarray}
  \Lambda_i^2 &\approx & \sqrt{ \frac{2}{b_i -4} K_i D[\alpha_i(M)]} M^2, \quad i = 1,2, \nn \\
  D[\alpha_i] & \approx & 0.1 \left(\frac{2\pi}{\alpha_i}\right)^6 e^{-\frac{2\pi}{\alpha_i}} ~ \equiv D_i, \nn \\
  K_i  &= &\prod_{j} \left(\frac{y_j}{4\pi}\right).
 \label{eq:ins}
\end{eqnarray}
$b_i$ is the beta function coefficients for $SU(3)_i$ above the scale $M$, for which one needs to take into account both the standard model fermions charged under $SU(3)_i$ and the colored link field. $D[\alpha]$ is the dimensionless instanton density~\cite{Callan:1977gz, Andrei:1978xg}. $K_i$ corresponds to the chiral suppression due to the light standard model quarks charged under $SU(3)_i$ and is a product over all the Yukawa couplings involved~\cite{Flynn:1987rs, Choi:1998ep}. In Ref.~\cite{Agrawal:2017ksf}, all the SM quarks are charged under $SU(3)_1$ so that $K_1 = \prod_{j=u, d, s, c, b, t} \left( y_j/(4\pi)\right) \approx 10^{-23}$, while $K_2 = 1$. More details of the derivation leading to the equations above could be found in Appendix~\ref{app:dilute}.
To compute the axion mass from the small instantons, one could run $\alpha_s$ from the weak scale to $M$ first. Combining it with Eq.~\eqref{eq:matching} and Eq.~\eqref{eq:ins}, we could compute $\Lambda_1, \Lambda_2$ as a function of $\alpha_1(M)$. The axion masses are then given by\footnote{Strictly speaking, this equation holds only when $f_a > M$. There could be further suppression due to additional fermionic states charged under $U(1)_{\rm PQ}$ with mass of order $f_a$ when $f_a < M$. Since we are only considering the case with $f_a \gg M$, we will not consider potential further suppression.}
\begin{equation}\label{eq:mi2}
  m_i^2 = \frac{\Lambda_i^2}{f_a},
\end{equation}
when $\Lambda_i \gg \Lambda_0$ in Eq.~\eqref{eq:IR}.
One could easily generalize the minimal model based on $SU(3)^2$ to a larger product gauge group $SU(3)^n$ with $n>2$. Then the matching condition of the gauge couplings at the symmetry breaking scale $M$ is
\begin{equation}
  \sum_{i=1}^n \frac{1}{\alpha_i(M)} = \frac{1}{\alpha_s (M)}.
\end{equation}
By enlarging the gauge group, one could increase the individual gauge coupling $\alpha_i$ and thus the corresponding contribution of the UV instantons. In this class of models, the regime we are mostly interested in is where all the axions are heavier than the standard QCD axion for a given $f_a$. Otherwise, if the lightest axion mostly obtains its mass from the IR instantons, it then behaves mostly like ordinary QCD axion.

\subsection{Dynamical Yukawa couplings}
\label{sec:Yukawa}
In our scenario, the small instantons are the dominant source for the axion potential during inflation. To relax the misalignment angle, we must have the masses of all the axions around or above the inflationary Hubble scale, $H_I$. This leads to a parametric inequality, for every axion (labelled by $i$),
\begin{equation}
  \sqrt{\tilde{K}_i} \frac{M^2}{f_a} \gtrsim \tilde{m}_i^\uv \gtrsim H_I  \Rightarrow M^2 \gtrsim H_I \frac{f_a}{\sqrt{\tilde{K}_i}}.
 \label{eq:lower}
\end{equation}
From Eq.~\eqref{eq:ins}, one could see that the inequality is only close to be saturated when the gauge coupling is large enough so that there is no significant exponential suppression from $e^{-2\pi/\alpha_i}$ in the instanton density $D_i$.

To have the usual slow roll inflation, we require that the link field energy density to be subdominant,
\begin{equation}
  \mV(\Sigma) \sim M^4 < H_I^2 M_\Pl^2 \Rightarrow M^2 < H_I M_\Pl,
\label{eq:higher}
\end{equation}
where $M_\Pl$ is the Planck scale and we take all the dimensionless couplings (the quartic couplings of the link fields and gauge couplings at scale $M$) to be of order one. Note that unlike the four general requirements discussed in Sec.~\ref{sec:mechanism}, this constraint on the symmetry breaking scale is specific to our model.\footnote{One may add a constant in the link field potential to cancel contribution from the remaining terms to evade the bound, which is clearly a fine tuning. While adding a constant term in the scalar potential is not unusual in the cosmology literature (\eg, in the hybrid inflation~\cite{Linde:1993cn}), we will not pursue it here.}
By combining Eq.~\eqref{eq:lower} and Eq.~\eqref{eq:higher}, we find that
\begin{equation}
  f_a < \sqrt{\tilde{K}_i } M_\Pl.
\label{eq:fbound}
\end{equation}
One could see an immediate challenge: if any of the light fermion suppression factors, $\tilde{K}_i$, is small during inflation, we could not raise the cosmological upper bound on $f_a$ to be above $10^{12}~\Gev$ at all! Indeed, in Ref.~\cite{Agrawal:2017ksf}, all the standard model quarks are charged under $SU(3)_1$ and thus $\tilde{K}_1 \sim 10^{-23}$. One could consider alternative constructions with different generations assigned to different $SU(3)$'s. For instance, we could consider a $SU(3)^3$ model with $u, d$ charged under $SU(3)_1$, $c, s$ charged under $SU(3)_2$ and $t, b$ charged under $SU(3)_3$. Then we have $\tilde{K}_1 \sim 1.5 \times 10^{-12}, \tilde{K}_2 \sim 2.5 \times 10^{-8}, \tilde{K}_3 \sim 1.5 \times 10^{-4}$. In this case, we could still not raise the cosmological bound on $f_a$ for the lightest axion given the tiny $\tilde{K}_1$. The only approach that allows us to evade the constraint is to raise the Yukawa couplings of the standard model quarks to be of order one during inflation to make all the $\tilde{K}$ factors larger. After inflation, the Yukawa couplings have to be relaxed to the standard model values and the UV instanton contribution to the lightest axion could then be suppressed by the corresponding small $K$ factor.

Now we present a concrete construction based on the discussions above and estimate the energy scales involved given all the constraints and requirements. Consider a $SU(3)^n$ model with all the standard model fermions charged under $SU(3)_1$. There will be a set of link fields $\Sigma_{12}, \Sigma_{23}, \Sigma_{34}, \cdots, \Sigma_{n-1,n}$, transforming as bi-fundamentals under the adjacent $SU(3)$'s and higgsing the product gauge group to the diagonal at the scale $M$. The $\beta$-function coefficient for each gauge coupling is
\begin{equation}
  b_1= \frac{13}{2}, ~ b_2=b_3 = \cdots = b_{n-1} = 10, ~ b_n = \frac{21}{2},
\end{equation}
which takes into account of the standard model fermions (only for $SU(3)_1$) and link fields (two for $SU(3)_i$ with $i= 2, \cdots, n-1$ and one for $SU(3)_1$ and $SU(3)_n$). The positive coefficients imply that all the gauge theories are asymptotically free. Consider that during inflation, all the Yukawa couplings are of order one and after the inflation, the Yukawa couplings return to the standard model values. We will elaborate the mechanism of dynamical Yukawa couplings later in this section.

Following \Eqs{eq:ins}{eq:mi2}, we can write the requirements for the masses of all the axions as:
\begin{eqnarray}
  \tilde{m}_1^\uv & \approx & 3 \times 10^{-4} ~ \frac{M^2}{f_a} \sqrt{D_1} ~ \sqrt{ \frac{\tilde{K}_1}{10^{-7}} }\nn\\
  & \sim & H_I \ , \label{eq:uv1}\\
  \tilde{m}_i^\uv & \approx & \frac{M^2}{f_a} \sqrt{D_i} \nn\\
  & > & H_I , \quad \quad \quad  i = 2,\cdots n , \label{eq:uvi}
\end{eqnarray}
where we have taken all the standard model Yukawa couplings to be $\sim \mO(1)$ during inflation, which gives $\tilde{K}_1 \sim 1/(4\pi)^6 \sim 10^{-7}$ due to the loop factors. Since there are no fermions charged under the other $SU(3)$'s, their associated $K_i$'s are all one. Note that we have demanded $\tilde{m}_1^\uv$ to be around $H_I$ and not much bigger. The reason for this will become clearer in \Sec{sec:relicabundance}. The gist is that for $\tilde{m}_1^\uv > H_I$ during inflation the misalignment angle is diluted exponentially with the number of {\it e}-folds of inflation, which results in an incredibly tiny value of the relic abundance.

As we discussed at the beginning of this section, the link field energy density has to be below the inflaton energy density $M^4 < 3 H_I^2 M_\Pl^2$, which we can rewrite as
\begin{equation}\label{eq:energy}
  M^2 < \bl( 4 \times 10^{18} ~ \Gev \br) ~ H_I .
\end{equation}

After inflation, due to the Yukawa suppression, we have for the lightest axion
\begin{eqnarray}\label{eq:irinst}
  m_1^\uv &=& \sqrt{y_u y_d y_c y_s y_t y_b} ~ \tilde{m}_1^\uv \approx 5.6 \times 10^{-9} ~ \tilde{m}_1^\uv <  m^\ir \nn\\
  &\Rightarrow &2 \times 10^{-12} ~ M^2 \sqrt{D_1} \sqrt{ \frac{\tilde{K}_1}{10^{-7}} } <  \Lambda_0^2 \approx (75.5 \; {\rm MeV})^2,
\end{eqnarray}
where $\Lambda_0$ characterize the usual contribution from the IR instantons. Note that the masses of the heavier axions do not change after inflation. Thus they remain heavy and are decoupled from the low energy EFT, which only contains the lightest axion, $a_1$.

\begin{figure}
\centering
\includegraphics[width=0.55\textwidth]{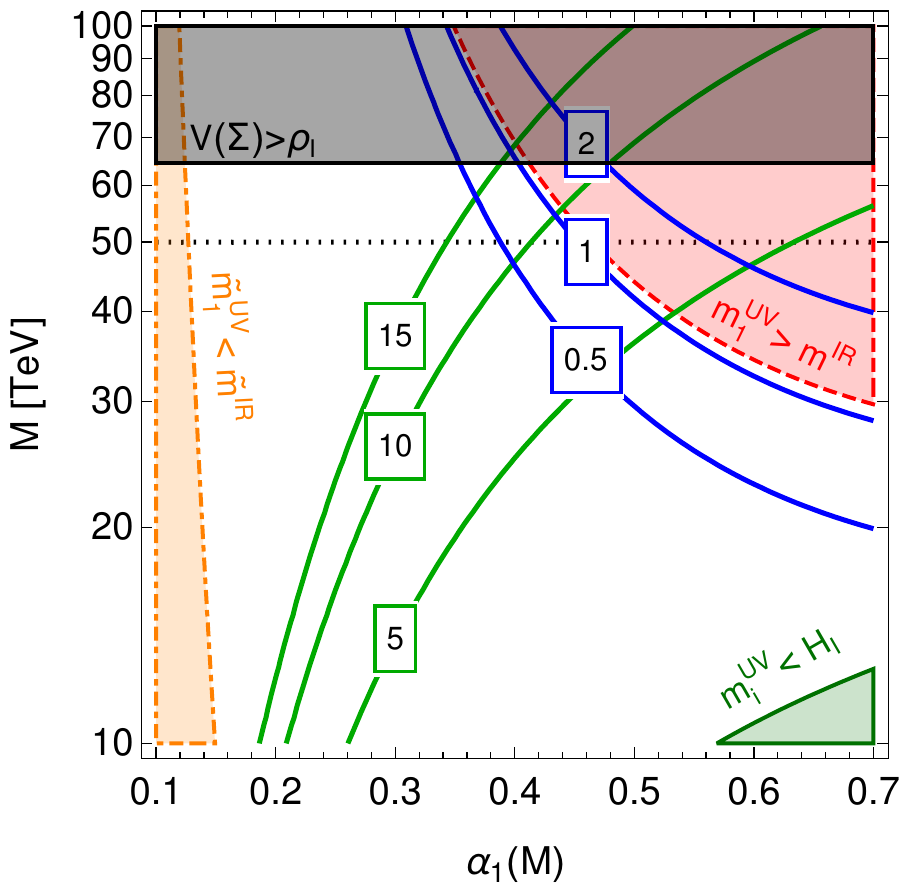}
\includegraphics[width=0.65\textwidth]{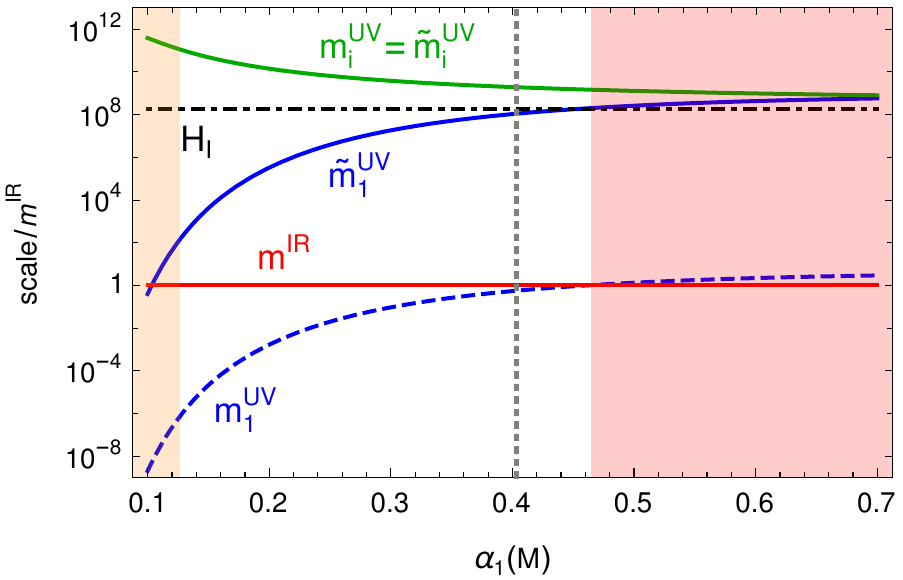}
\caption{Results of $SU(3)^4$ model, with PQ scale $f_a = 10^{15}~\Gev$, inflation scale $H_I = 1~\ev$, and chiral suppression $\tilde{K}_1=3\times10^{-7}$ during inflation. We fix $\alpha_2(M) = \alpha_3(M) = \alpha_4(M)$.  {\it Top:} Allowed $\alpha_1(M) - M$ parameter space, after excluding those areas that do not satisfy the requirements of the model. The blue and green lines are are the curves of fixed $\tilde{m}_1^\uv/H_I$ and $m_i^\uv/H_I$ respectively. {\it Bottom:} The ratio of different mass/energy scales to the QCD axion mass $m^\ir$, with $M=50~\Tev$, which corresponds to the black dotted line of the top plot. The red and orange shaded areas are the same to those in the top plot: regions with $m_1^\uv > m^\ir$ and $\tilde{m}_1 < \tilde{m}^\ir$ respectively. The vertical dotted gray line corresponds to the $\tilde{m}_1^\uv/H_I = 0.6$ benchmark, for which the $a_1$ axion's relic abundance is equal to the observed abundance of dark matter today. Note that the UV mass for axion with $i \neq 1$ (green line) does not change during and after inflation.}
\label{fig:axionmass}
\end{figure}

The inequalities above combined tell us the relevant scales could be
\begin{eqnarray}
  M & \lesssim & 50 ~ \Tev ~ D_1^{-1/4} ~ \bl( \frac{\tilde{K}_1}{10^{-7}} \br)^{-1/4} \ ,\label{eq:mscale} \\
  f_a & \lesssim & 10^{15} ~ \Gev ~ \sqrt{D_1} ~ \bl( \frac{\tilde{K}_1}{10^{-7}} \br)^{1/2} \ ,\label{eq:fscale} \\
 \ev ~ \left(\frac{M}{50~\Tev}\right)^2 \lesssim H_I  & \lesssim & 2.5 \times 10^3  \ev ~ \left(\frac{M}{50~\Tev}\right)^2 \left(\frac{10^{15} \, {\rm GeV}}{f_a}\right). \label{eq:hscale}
\end{eqnarray}
Note that we can increase the upper bound on $f_a$ with either a larger instanton density $D_1$ or a larger $\tilde{K}_1$ prefactor. On one hand, $\alpha_1 \sim 0.5 \Rightarrow D_1 \sim \mO(1)$ and larger densities require couplings in the non-perturbative regime, where our estimates are no longer valid. On the other hand, larger $\tilde{K}_1$ requires the Yukawas during inflation to be larger than $\sim \mO(1)$. Thus we could not raise $f_a$ to be much above $10^{15}~\Gev$ in the perturbative regime of our scenario. A more precise result for the allowed parameter space can be found in Fig.~\ref{fig:axionmass}, where we consider an $SU(3)^4$ model with axion constant $f_a = 10^{15}~\Gev$, inflation scale $H_I = 1~\ev$, and chiral suppression factor $\tilde{K}_1 = 3 \times 10^{-7}$ with all the Yukawa couplings taken to be one during inflation.

There is a subtlety regarding the early IR instanton contribution to the axion potential. Initially, since during inflation the SM fermions have all bigger masses (because $y_f \sim \mO(1)$), the running of $\alpha_s$ changes and so does the QCD scale (which we take to be the scale at which $\alpha_s(M_\QCD) = 4 \pi$). Indeed, while after inflation we have the SM value $M_\QCD \sim 100~\Mev$, during inflation we have $\tilde{M}_\QCD \sim 900~\Mev$. This implies that the IR instanton contribution during inflation, $\tilde{m}^\ir$, is enhanced by $\left(\tilde{M}_\QCD/M_\QCD\right)^2 \sim 100$, which is still subdominant when compared to the UV contributions given in \Eqs{eq:uv1}{eq:uvi} with the scales from \Eqst{eq:mscale}{eq:hscale}.

Having found the allowed parameter space for our model, we now present a mechanism to vary the Yukawa couplings. A similar mechanism has been used in the literature to achieve a strong first-order electroweak phase transition~\cite{Baldes:2016gaf}. Consider a Froggatt-Nielsen (FN) model~\cite{Froggatt:1978nt} with a flavon $S$, which is a complex scalar charged under a $U(1)_{\rm FN}$. The potential of $S$ is
\begin{equation}\label{eq:vs}
  \mV(S) = (- \mu_s^2 - g \phi^2) S^\dagger S + \lambda_s (S^\dagger S)^2 - A^2 (SS + S^\dagger S^\dagger),
\end{equation}
where $\phi$ is the inflaton. All the parameters are real.\footnote{In particular, it is important that $A$ is real. This could be realized if CP is a gauge symmetry broken spontaneously only in the sector that generates the $y$'s.} Note that the last term explicitly breaks the FN symmetry which the flavon is charged under and forces the VEV of the scalar to be real. The VEV squared of $S$ is
\begin{equation}
  2 \langle S \rangle^2 \equiv v_s^2 = \frac{\mu_s^2+ g \phi^2 + 2 A^2}{\lambda_s}.
\end{equation}
$S$ couples to the standard model quarks as
\begin{equation}\label{eq:yukawaterms}
  \mL = y^u_{ij} \left(\frac{S}{\Lambda_s}\right)^{m_{ij}} \bar{Q}_i \tilde{H} u_j + y^d_{ij} \left(\frac{S}{\Lambda_s}\right)^{n_{ij}} \bar{Q}_i H d_j,
\end{equation}
where $y$'s are dimensionless order one numbers; $i,j$ indicate the generations and $m, n$ indicate the power dependence of the flavon VEV, which depends on the FN charges of the fields involved. For instance, we could have the following charge assignment for $S$ and the three generations of quarks:
\begin{equation}
S: -1, \, \bar{Q}_3: 0, \, \bar{Q}_2: 2, \, \bar{Q}_1: 3, \, u_3: 0, \, u_2: 1, \, u_1: 4, \, d_3: 2, \, d_2: 2, \, d_1: 3.
\label{eq:charge}
\end{equation}
Then the up and down Yukawa matrices, in the flavor basis, are given by
\[
y_u=
  \begin{pmatrix}
    y^u_{11} \epsilon^7 & y^u_{12} \epsilon^4 & y^u_{13} \epsilon^3 \\
   y^u_{21} \epsilon^6 &y^u_{22} \epsilon^3 &y^u_{23} \epsilon^2 \\
   y^u_{31} \epsilon^4 & y^u_{32} \epsilon &y^u_{33} \\
  \end{pmatrix}
\]
and
\[
y_d=
  \begin{pmatrix}
    y^d_{11} \epsilon^6 & y^d_{12} \epsilon^5 & y^d_{13} \epsilon^5 \\
   y^d_{21} \epsilon^5 &y^d_{22} \epsilon^4 &y^d_{23} \epsilon^4 \\
   y^d_{31} \epsilon^3 & y^d_{32} \epsilon^2 &y^d_{33} \epsilon^2 \\
  \end{pmatrix}
\]
where
\begin{equation}\label{eq:epsilon}
  \epsilon = \frac{v_s}{\sqrt{2}\Lambda_s}.
\end{equation}
For $\epsilon \sim 0.2$, one could get the standard model quark masses and mixing today. For $\epsilon \sim 1$, one could get order one Yukawa couplings and mixing. During inflation, $g \phi^2$ could be large and the VEV of $S$ could be increased by one order of magnitude. After inflation, when the inflaton field value decreases, $\epsilon$ reduces to 0.2 and the Yukawa matrices are settled to the values today.

Analogous to \Eq{eq:higher}, in order for inflation to take place, the flavon potential must be subdominant:
\begin{equation}
  \mV(S ) \sim \frac{\lambda_s v_s^4}{4} = \lambda_s \epsilon^4  \Lambda_s^4 < 3 M_\Pl^2 H_I^2 \ ,
\end{equation}
which for $\lambda_s \sim \mO(1)$ and $\tilde{\epsilon} \sim 1$ immediately translates into the bound:
\begin{equation}\label{eq:lsbound}
  \Lambda_s < 60~\Tev ~ \bl( \frac{H_I}{\ev} \br)^{1/4} \ .
\end{equation}

\begin{figure}
\centering
\includegraphics[width=0.7\textwidth]{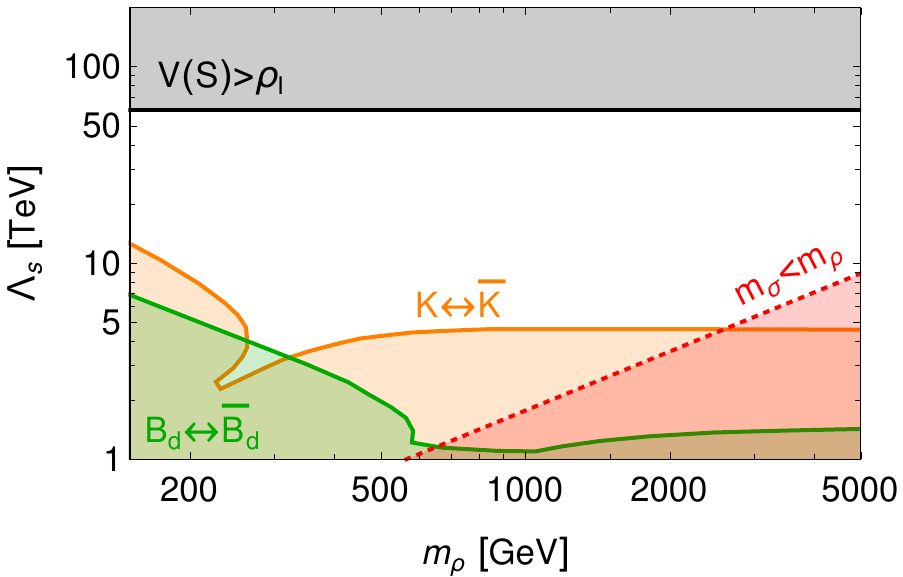}
\caption{Allowed $m_\rho-\Lambda_s$ parameter space for the FN model. In orange and green are the flavor bounds from $K \leftrightarrow \overline{K}$ and $B_d \leftrightarrow \overline{B}_d$ mixings \cite{Bona:2007vi}, taken from Ref.~\cite{Bauer:2016rxs}. The black region does not satisfy the inflation condition \Eq{eq:lsbound}, and the red, dotted region violates the consistency condition $m_\sigma > m_\rho$.}
\label{fig:flavon}
\end{figure}

Furthermore, from \Eq{eq:vs}, the real and imaginary parts of the flavon, which we respectively denote by $\sigma$ and $\rho$, acquire masses
\begin{eqnarray}\label{eq:fnmasses}
  m_\sigma^2 & = & 2\lambda_s v_s^2 = 4 \epsilon^2 \lambda_s \Lambda_s^2 = 2\mu_s^2 + 4A^2 + 2g \phi^2 \ , \\
  m_\rho^2 & = & 4 A^2 < m_\sigma^2 \ .
\end{eqnarray}
These scalars produce flavor-changing neutral currents \cite{Baldes:2016gaf, Bauer:2016rxs, Buras:2013rqa, Crivellin:2013wna}, which are tightly bound by meson anti-meson mixing. In Fig.~\ref{fig:flavon} we summarize these constraints, which we take from \cite{Bauer:2016rxs}, with $\lambda_s = 2$ as a benchmark. We have converted the published bounds on $v_s$ into bounds on $\Lambda_s$ by using \Eq{eq:epsilon} with $\epsilon = 0.2$. The dominant flavor constraints are $K \leftrightarrow \overline{K}$ and $B_d \leftrightarrow \overline{B}_d$ \cite{Bona:2007vi} oscillations. Also included are the bound from \Eq{eq:lsbound} and the consistency condition $m_\sigma > m_\rho$.

It is important that in this setup, the contribution to the strong CP phase from the quark mass matrix, $\mathrm{arg}(\mathrm{Det}M)$ with $M$ the quark mass matrix, does not change when the VEV of the flavon changes. In terms of the Yukawa matrices, $\mathrm{arg}(\mathrm{Det}M) = \mathrm{arg}\left[\left(\mathrm{Det} \,y_u\right)\left( \mathrm{Det} y_d \right)\right]$. Given the FN charge assignment in Eq.~\eqref{eq:charge}, one could find that
\begin{eqnarray}
  && \left(\mathrm{Det} \, y_u \right)  \left(\mathrm{Det} \, y_d \right) \propto  \epsilon^\alpha \epsilon^\beta  =\epsilon^{22}, \nn \\
  && \alpha=\sum_{i=1}^3 \left[q(Q_i) + q(u_i)\right] = 10, \quad \beta=\sum_{i=1}^3 \left[q(Q_i) + q(d_i) \right]= 12,
\end{eqnarray}
where $q$ indicates the FN charge of the quark. Note that the determinant of the Yukawa matrices is always proportional to a power of $\epsilon$, independent of the FN charge assignment, while the specific power is model dependent. Thus varying the VEV of the flavon doesn't change the strong CP phase from the quark mass matrix, provided that $\epsilon$ is real, and high dimensional operators such as $(S^\dagger S) S^{m_{ij}} \bar{Q}_i \tilde{H} u_j$ are suppressed.

There still exists a non-negligible radiative correction to the strong CP phase during inflation, which arises from order one CKM phase. The radiative corrections to $\bar{\theta}$ are minuscule in the standard model. The largest contribution comes from finite four-loop Cheburashka diagrams~\cite{Khriplovich:1985jr,Hiller:2002um},
\begin{equation}
  \bar{\theta}_\SM^{\rm finite} = -\frac{7}{9} \frac{\alpha_s}{4\pi}\left( \frac{\alpha_W}{4\pi}\right)^2 \frac{m_s^2 m_c^2}{m_W^4} J  \, \ln\frac{m_t^2}{m_b^2} \ln^2 \frac{m_b^2}{m_c^2} \left(\ln \frac{m_c^2}{m_s^2} + \frac{2}{3} \ln\frac{m_b^2}{m_c^2}\right),
\label{eq:radiative}
\end{equation}
where $\alpha_W$ is the weak coupling strength and $J$ is the Jarlskog invariant, given by:
\begin{equation}
  J \equiv \mathrm{Im} \bl( V_{us} V_{cb} V_{ub}^* V_{cs}^* \br) = s_{12}s_{13}s_{23}c_{12}c_{13}^2c_{23}\sin \delta \ ,
\end{equation}
where $V$ is the CKM matrix, $s_{ij} = \sin \phi_{ij}, c_{ij} = \cos \phi_{ij}$; and $\phi_{ij}$ and $\delta$ are the angles and phase of $V$ in the PDG parameterization.
For the standard model values of quark masses and mixings, this gives $\bar{\theta}_\SM^{\rm finite} \sim 10^{-19}$. There is also a logarithmic divergent contribution first arising from seven-loop diagrams, but it is even much smaller~\cite{Ellis:1978hq}.

For the assignments in \Eq{eq:charge}, the FN model gives \cite{Baldes:2016gaf}
\begin{equation}
  \vert V_{us} \vert \sim \vert V_{cd} \vert \sim \epsilon \ , \quad \vert V_{cb} \vert \sim \vert V_{ts} \vert \sim \epsilon^2 \ , \quad \vert V_{ub} \vert \sim \vert V_{td} \vert \sim \epsilon^3 \ ,
\end{equation}
which means that for our model $J \sim \mO(1)$ during inflation, when $\tilde{\epsilon} \sim 1$. Furthermore, the masses of the standard model quarks, which are only charged under $SU(3)_1$, are about the weak scale. Therefore the radiative correction to $\bar{\theta}_1$ is enhanced significantly, compared to that in the standard model. This leads to a mismatch between the minima of the QCD axion potentials during and after the inflation. Barring accidental mass degeneracies between the quarks, we estimate the mismatching angle to be
\begin{equation}
  \theta_\mis \sim \bl( \frac{\alpha_W}{4\pi} \br)^2 \sim 10^{-5},
\end{equation}
where in Eq.~\eqref{eq:radiative} we take  $\alpha_s/(4\pi) \sim 1$ around the GeV scale, ignore all order one numbers, and set all the mass ratios, their logarithms, and $J$, to order one.

\subsection{Axion spectrum and relic abundance}
\label{sec:relicabundance}
In our model, we have multiple axions. Taking the $SU(3)^4$ model as an example, we have four: the lightest axion, $a_1$, is the QCD axion, receiving most of its mass from the infrared instantons after inflation; while $a_2, a_3, a_4$ are significantly heavier, with their masses unchanged and dominated by the contributions from the UV instantons. In this section, we compute their relic abundances today.

First we estimate the diluted misalignment angle for each axion at the end of the inflation, measured from the minimum of the axion potential during inflation. The evolution of each axion during inflation is dictated by the equation of motion
\begin{equation}
  \ddot{a}_i + 3 H_I \dot{a}_i + V^\prime(a_i) = 0, \quad i=1,2,3,4.
\end{equation}
During inflation, the Hubble scale is approximately a constant, and we will approximate $V(a_i) \approx \tilde{m}_i^2 a_i^2/2$ so that $V^\prime (a_i) = \tilde{m}_i^2 a_i$. Here the axion mass is from the UV instantons, $\tilde{m}_i = \tilde{m}_i^\uv$. Changing variable from time to the number of $e$-folds by using $dN=H_I dt$, we have
\begin{equation}
  \frac{d^2 a_i}{d N^2} + 3 \frac{d a_i}{d N} + \frac{\tilde{m}_i^2}{H_I^2} a_i =0.
\end{equation}
There are two interesting limits where we could obtain analytical solutions:
\begin{itemize}
  \item $\tilde{m}_i \gg H_I$: the $i$-th axion oscillates during inflation and the energy density redshifts as matter: $\rho_i = \tilde{m}_i^2 a_i^2 \propto R^{-3}$, with $R$ the scale factor. Thus $\langle a_i \rangle = a_{0;i} ~ e^{-3N/2}$, where $\langle a_i \rangle$ is the cycle average of the axion field value (or in other words, the oscillation amplitude) and $a_{0;i} \sim f_a$ is the initial amplitude without tuning. So at the end of inflation, $\theta_{\dil;i} = \theta_{0;i} ~ e^{-3N/2}$, and the initial angle $\theta_{0;i}$ is generically of order one.
  \item $\tilde{m}_i < H_I$: the axion rolls down its potential without oscillation. Solving the classical equation of motion gives its displacement at the end of the inflation as $\theta_{\dil;i} =\theta_{0;i} ~ e^{-N m_i^2/(3H_I^2)}$. The axion amplitude redshifts away more slowly, compared to the first case.
\end{itemize}

From Fig.~\ref{fig:axionmass} we can see that in the allowed region of parameter space, $a_{2,3,4}$ could be much heavier than $a_1$ (green contours), while the mass of $a_1$ can be close to the inflationary Hubble scale (blue contours). Thus $a_{2,3,4}$ belong to the first limit with a significantly exponentially suppressed misalignment angle $\theta_{\dil;i}$; while $a_1$ is in the second limit with a small but non-negligible $\theta_{\dil;1}$.

Assuming a standard post-inflationary history (inflation$\rightarrow$reheating (rh)$\rightarrow$radiation domination), the number of {\it e}-folds of inflation, $N$ must satisfy the condition:
\begin{eqnarray}
  && \frac{1}{H_I} e^{N} \frac{R_\mathrm{rh}}{R_\mathrm{end}} \frac{T_\mathrm{rh}}{T_0} \bl( \frac{g_*(T_\mathrm{rh})}{g_*(T_0)} \br)^\frac{1}{3} \geq \frac{1}{H_0} \ , \nn\\
  \Rightarrow && N \geq 35 + \log \bl[ \bl( \frac{H_I}{\ev} \br) \bl( \frac{50 ~ \Tev}{T_\mathrm{rh}} \br) \bl( \frac{R_\mathrm{end}}{R_\mathrm{rh}} \br) \bl( \frac{g_*(T_\mathrm{rh})}{g_*(T_0)} \br)^\frac{1}{3} \br] \ ,
\end{eqnarray}
where $H_0$ is the Hubble expansion rate today, and $g_*(T_\mathrm{rh})$ and $g_*(T_0)$ are the effective degrees of freedom for entropy at reheating and today. Note that if reheating happens shortly after the end of inflation, $R_\mathrm{rh} \approx R_\mathrm{end}$ and $T_\mathrm{rh} \approx \sqrt{H_I M_\Pl} \approx 50~\Tev \bl( \frac{H_I}{\ev} \br)^{1/2}$. Given this high reheating temperature, the heavy gauge bosons could be produced, which are unstable and decay to the standard model quarks.

Taking the benchmark scales from \Eqst{eq:mscale}{eq:hscale}, for the $SU(3)^4$ model with $\alpha_2=\alpha_3=\alpha_4$, $\tilde{K}_1 = 3 \times 10^{-7}$, and $N=35$ {\it e}-folds; we find
\begin{eqnarray}
  &&\tilde{m}_1^\uv/H_I = 0.59 \ , \quad \tilde{m}_i^\uv/H_I = 11  \nn \\
  \Rightarrow && \theta_{\dil;1} = 0.017 ~ \theta_{0;1} \ , \quad \theta_{\dil;i} =  10^{-17} \theta_{0;i} \ ,  \quad i = 2,3,4.
 \label{eq:misalignment}
\end{eqnarray}

After inflation, $a_1$ becomes much lighter and behaves as the ordinary QCD axion. The initial misalignment angle of its oscillation when the Hubble scale drops around its mass is $\theta_1 = {\rm max} [\theta_{\dil;1}, \theta_\mis]$ as discussed in Sec.~\ref{sec:mechanism}. The potentials of the other heavy axions, $a_{2,3,4}$ do not change. They continue oscillating after the end of inflation, with the amplitude reducing as $R^{-3/2}$. The initial diluted misalignment angles for their oscillations after inflation (measured from the minima of the potentials during inflation) are tiny, as shown in Eq.~\eqref{eq:misalignment}. Note that, since the standard model quarks are only charged under $SU(3)_1$, then the radiative correction from the CKM phase only contributes to the strong phase of $SU(3)_1$. The displacement between the minima of the heavy axion's potentials during and after inflation is only generated through the small mass mixing induced by QCD confinement between the heavy axion and the lightest axion $a_1$, which is also highly suppressed by the square of the mass ratios.\footnote{We assume that there is no mixing between different axions due to cross couplings such as $a_1 G_2 \tilde{G}_2$.} Therefore we could safely ignore their relic abundances today. The relic abundance of the QCD axion is found to be
\begin{equation}
  \Omega_1 h^2 \approx 0.12 \, \left(\frac{\theta_1}{0.017}\right)^2 \left(\frac{f_a}{10^{15} \, \Gev}\right)^{\frac{n+6}{n+4}} \left(3.8 \right)^\frac{2}{4+n} \left(10\right)^{\frac{4-n}{4+n}} \left(\frac{\chi_0(1.5)}{3.7 \times 10^{-14} \, \Gev^4}\right)^{-\frac{1}{4+n}}.
\label{eq:abundance}
\end{equation}
We adopt the equation from Ref.~\cite{Dine:2017swf}. $n$ is in the range of $7-20$, which is the power dependence of the vacuum energy on temperature; while $\chi_0(1.5)$ is the topological susceptibility at $T=1.5~\Gev$. The benchmark from \Eq{eq:misalignment} makes the relic abundance of the QCD axion equal to that of the DM by taking $n=8$ in \Eq{eq:abundance}; it is shown as the vertical dotted gray line in the bottom plot of \Fig{fig:axionmass}.

\section{Other possibility and challenges}
\label{sec:other}
The ability of our model to increase the UV instanton contribution to the axion mass during inflation for $f_a > 10^{12}~\Gev$ hinges on the dynamical Yukawa mechanism. Indeed, it is only by having the Yukawas be $\sim \mO(1)$ during inflation that the chiral suppression to the instanton, $\tilde{K}_1$, becomes large enough to accommodate the necessary masses.

A different way to vary the axion mass during and after inflation is to dynamically change the symmetry breaking scale. The idea is to take advantage of the fact that the instanton density, $D_1[\alpha_1(M)]$, is exponentially sensitive to the gauge coupling. If the gauge symmetry breaking scale $M$ is {\it higher} after inflation than during inflation, $\alpha_1(M)$ decreases because of its RG running, and therefore so does $D_1[\alpha_1(M)]$. Then the axion mass is suppressed after inflation. This change in the symmetry breaking scale can be achieved by coupling the link fields $\Sigma$'s that break the gauge symmetries to the inflaton dynamics, analogous to how we change the VEV of the flavon $S$ after inflation.

However, this alternative setting is plagued with several problems. The most important one is that, if there is no dynamical Yukawa mechanism, the chiral suppression remains constant and there is always at least one small $K$ factor no matter how different generations of standard model quarks are coupled to different gauge groups. In addition, the energy density in the link fields must be subdominant during inflation and at most of order $H_I^2 M_\Pl^2$ after inflation. These requirements, which are encoded in Eq.~\eqref{eq:lower} and~\eqref{eq:higher} plus an additional equation $\mV(\Sigma)_\aft \lesssim H_I^2 M_\Pl^2$ with $\mV(\Sigma)_\aft$ the link field potential after inflation, restricts the separation of the symmetry-breaking scales $M$ during and after inflation (more concretely, the separation could be only a factor of $\sim 100$). As a result, there is simply not enough RG ``running distance'' for $D_1[\alpha_1(M)]$ to make $m_1^\uv$ large during inflation and small after it. Of course, this setting can be used in conjunction with the FN dynamical Yukawas, but as we have shown in this paper the latter is sufficient to satisfy all the requirements by itself.

Lastly we comment on several challenges in our scenario. From Eq.~\eqref{eq:hscale} one could see that our scenario works with very low-scale inflation, which is not easy to achieve in general. There is no no-go theorem though. We will not attempt it in this article and just want to point out that there exist constructions based on the inflaton as an axion-like particle in Refs.~\cite{Daido:2017wwb, Daido:2017tbr} with Hubble scale around or below eV. In addition, there are two coincidences in energy scales. Unlike previous models that relax the misalignment angle dynamically, we do not assume the angle to be of order $10^{-2} - 10^{-3}$ due to new CP violating phases. We compute and demonstrate that the mismatching between the minima of the axion potentials during and after inflation is tiny. Then to have the misalignment angle to be about $10^{-2}$, we cannot have too much dilution of the axion oscillation amplitude during inflation. This leads to the requirement that the mass of the lightest axion is close to the Hubble scale during inflation. Another coincidence resides in the dynamical Yukawa coupling mechanism. Since the FN symmetry breaking spurion $\epsilon$ only varies by a factor of a few during and after inflation, we need the contribution to the flavon mass squared from its coupling to the inflaton during inflation to be within one order of magnitude above its mass squared parameter. In addition, we have not tried to construct a full UV completed flavor model to generate the ${\cal O}(1)$ CKM phase through phases in the dimensionless coefficients $y$'s in the FN model and explain the realness of the $A$ parameter in the flavon potential. We leave it for future work to realize these two coincidences and construct a full-fledged flavor model in a fully dynamical way.

\section{Conclusions and outlook}
\label{sec:conclusion}

Many current-generation and proposed next-generation experiments are searching for the low-mass QCD axion and/or axion-like particles.
In particular, CASPEr \cite{Graham:2013gfa, Budker:2013hfa} and ABRACADABRA \cite{Kahn:2016aff} might be able to reach the QCD axion line for values of the decay constant $f_a$ corresponding to the GUT scale and beyond. In order for these experiments to detect a positive signal, the axion needs to be a significant fraction of the dark matter in the Universe. It is not easy to have a consistent cosmological history with such large values of $f_a$. Indeed, to avoid overclosing the Universe, the cosmological bound is $f_a \lesssim 10^{12}~\Gev$ for models with an initial axion misalignment angle $\theta_0$ of $\mO(1)$.

In this paper we have constructed a model to dynamically suppress the QCD axion misalignment angle and thereby relax the cosmological bound on $f_a$. We achieve this by raising the contributions to the QCD axion's mass coming from small UV instantons during inflation, and subsequently lowering them once the inflationary era ends. To make the UV instanton change in this way we require two main ingredients: ({\it i}) making the color group $SU(3)_c$ the result of higgsing a product gauge group $SU(3)^n$ at a high energy scale, and ({\it ii}) allowing dynamical Yukawas for the standard model quarks with a FN mechanism, whose flavon has a VEV that changes dynamically via its coupling to the inflaton. Compared to early work on dynamical misalignment, we study in more details the CP violation in the model and address the key question whether the minimum of the QCD axion potential shifts at early and late times.

Different consistency and experimental bounds restrict the parameters in both the symmetry breaking and the flavor structure ingredients, resulting in a scenario with specific values for the different scales involved. In particular, we require an inflationary scale around $1~\ev$, both the symmetry breaking and FN scales to be of order $\sim \mathrm{few}\times 10~\Tev$, and a maximum value of $f_a \sim 10^{15}~\Gev$.
The new particles associated with the gauge symmetry breaking and FN scales could be within reach of the future generation experiments, either from direct production at a future hadron collider or from indirect flavor measurements. This suggests an interesting and exciting possibility to have simultaneous discoveries at different experimental frontiers: low-energy axion detection and high-energy collider experiments!


\section*{Acknowledgments}
We thank Prateek Agrawal, Raymond Co, Lisa Randall, Matt Reece and Martin Schmaltz for useful discussions and comments. The project was initiated at KITP in UCSB, which is supported by the National Science Foundation under Grant No. NSF PHY-1748958. MBA and JF are supported by the DOE grant DE-SC-0010010 and NASA grant 80NSSC18K1010.

\appendix

\section{Estimating contribution of UV instantons to the axion potential}
\label{app:dilute}

In this appendix, we provide more details of the estimates of UV instantons that lead to Eqs.~\eqref{eq:ins}. The scale of the axion potential from the UV instanton is given by
\beq
\Lambda_i^4= \int_{\rho=0}^{\rho=1/M} 2 \frac{d\rho}{\rho^5} D[\alpha_i(1/\rho)],
\eeq
where the instanton density $D$ is given by the second line in Eqs.~\eqref{eq:ins}. The one-loop running gauge coupling is governed by
\beq
\frac{d \alpha_i^{-1}}{d \ln \mu} = \frac{b_i}{2\pi} \ ,  \ \Rightarrow  e^{-\frac{2\pi}{\alpha_i(1/\rho)}} = e^{-\frac{2\pi}{\alpha_i(1/M)}} (\rho M)^{b_i}.
\eeq
The instanton density could be approximated as
\beq
D[\alpha_i (1/\rho)] \approx D[\alpha_i (M)] (\rho M)^{b_i}.
\eeq
Plugging the approximation above to the integration, we find that the contribution is dominated by instantons with size $\rho \sim 1/M$ and obtain Eqs.~\eqref{eq:ins}.

\bibliography{ref}
\bibliographystyle{utphys}
\end{document}